\documentclass{PoS}
\usepackage{gensymb}
\usepackage{graphicx}
\usepackage{subfig}
\usepackage{cite}
\usepackage[utf8]{inputenc}
\DeclareGraphicsExtensions{.pdf,.jpeg,.png}

\title{A Search for Cosmic-ray Proton Anisotropy with the Fermi Large Area Telescope}

\ShortTitle{Fermi-LAT Proton Anisotropy}

\author{\speaker{Matthew Meehan}, Justin Vandenbroucke on behalf of the Fermi-LAT Collaboration\\
        Department of Physics and Wisconsin IceCube Particle Astrophysics Center, University of Wisconsin, Madison, WI 53706, USA\\
        E-mail: \email{mrmeehan@wisc.edu}}

\abstract{In eight years of operation, the Fermi Large Area Telescope (LAT) has detected a large sample of cosmic-ray protons. The LAT's wide field of view and full-sky coverage make it an excellent instrument for studying anisotropy in the arrival directions of protons at all angular scales. These capabilities enable the LAT to make a full-sky 2D measurement of cosmic-ray proton anisotropy complementary to many recent TeV measurements, which are only sensitive to the right ascension component of the anisotropy. Any detected anisotropy probes the structure of the local interstellar magnetic field or could indicate the presence of a nearby source. We present the first results from the Fermi-LAT Collaboration on the full-sky angular power spectrum of protons from approximately 100 GeV - 10 TeV.}

\FullConference{35th International Cosmic Ray Conference --- ICRC2017\\
		10--20 July, 2017\\
		Bexco, Busan, Korea}

\begin{document}

\section{Introduction} \label{intro}
Anisotropy in the arrival directions of cosmic rays has been detected by many ground-based observatories over the past several decades~\cite{nagashima, hall, tibet_2005, icecube_anisotropy}. A large-scale (dipole) anisotropy is consistently seen with an amplitude of $10^{-4}-10^{-3}$ at energies from hundreds of GeV to \textasciitilde5 PeV. The dipole amplitude and phase both exhibit an energy dependence which has recently been resolved up to PeV energies~\cite{icecube_anisotropy, tibet_2017}. There are a number of possible explanations for the observed dipole: a nearby source of cosmic rays which dominates the large-scale diffusion gradient or the influence of a strong, local magnetic field are both expected to create a dipole anisotropy~\cite{ibex}. It has recently been shown that a combination of these effects can explain the observed dipole behavior~\cite{ deciphering_dipole}. A dipole due to the relative motion of the solar system through the isotropic cosmic-ray plasma known as the Compton-Getting effect is also expected, though none has been detected to date~\cite{compton_getting}. Small-scale anisotropy at the level of $10^{-5}-10^{-4}$ has been seen at angular scales as small as 10\degree~\cite{hawc}, which may be due to the scattering of cosmic rays off of local turbulent magnetic fields~\cite{ahlers_turbulence, vanessa_turbulence}.

In order to measure the anisotropy at these small amplitudes, data-driven methods are used to estimate the exposure, which needs to be known to an accuracy of < $10^{-3}$. When using these data-driven analysis methods, ground-based observatories lose all sensitivity to the declination component of the anisotropy~\cite{deciphering_dipole}. The measured anisotropy is thus a projection onto right ascension. The Fermi Large Area Telescope (LAT) is sensitive to anisotropy in both right ascension and declination, adding new information to the cosmic-ray anisotropy mystery. Furthermore, it sees the entire sky, which is not possible with any single ground-based instrument.

\section{Fermi Large Area Telescope} \label{instrument}
The Fermi Large Area Telescope (LAT) is a pair-conversion gamma-ray telescope onboard the Fermi Gamma-ray Space Telescope. It was launched in June of 2008, entered full data taking operations in August 2008, and has accumulated > 8 years of data. The LAT is optimized to detect gamma rays via the electron-positron pairs created when they interact with nuclei inside the detector. Because it is essentially a charged-particle detector, it has the ability to study other charged cosmic rays such as protons, electrons, and positrons~\cite{fermi_positron, fermi_cre_spectrum, fermi_cre_anisotropy, fermi_cre_anisotropy_2010}.

The LAT is a survey instrument with a wide, 2.4 sr instantaneous field of view that surveys the entire sky every 2 orbits or \textasciitilde3 hours. It is able to see the entire sky by rocking N/S from zenith on each successive orbit. This rocking angle was 35\degree{} in the first 13 months of science operations and increased to 50\degree{} thereafter~\cite{fermi_pass7}. The increased rocking angle allows the LAT to explore larger zenith angles, but also introduces more background from the Earth limb. This is taken into account in the following sections to avoid contamination of the signal due to geomagnetic deflection.

There are three main sub-systems in the LAT: the tracker (TKR), calorimeter (CAL), and anti-coincidence detector (ACD). The tracker is composed of 18 layers of alternating x and y silicon strips for direction reconstruction and tungsten foils to promote the conversion of gamma rays to e+/e-. The calorimeter consists of eight layers of CsI crystals in a hodoscopic arrangement which allows the LAT to measure a 3D profile of each shower. The anti-coincidence detector consists of segmented scintillator panels wrapped around the outside of the detector to identify charged particles~\cite{fermi_pass7}.

\section{Event Selection} 
\label{dataset}
The data set used in this analysis contains 160 million events recorded between December 2008 and December 2016. We developed a custom event selection in order to construct a proton data set suitable for an anisotropy search. The event selection begins with the same selection used to measure the Fermi-LAT proton spectrum. A set of minimum quality cuts is imposed first to select events with accurately reconstructed variables. All events must have a track in the tracker, traverse a path length of at least four radiation lengths in the calorimeter, and deposit > 20 GeV of energy in the calorimeter. The onboard filter accepts all events satisfying this last requirement, which ensures that the filter efficiently selects high-energy protons. Additionally, there are two minimum quality requirements on the direction reconstruction; one to select events with good angular resolution and another to reject back-entering events.

\begin{figure}
	\captionsetup[subfigure]{labelformat=empty}
 	\centering
	\subfloat[]{\includegraphics[width=7cm, height=4.87cm]{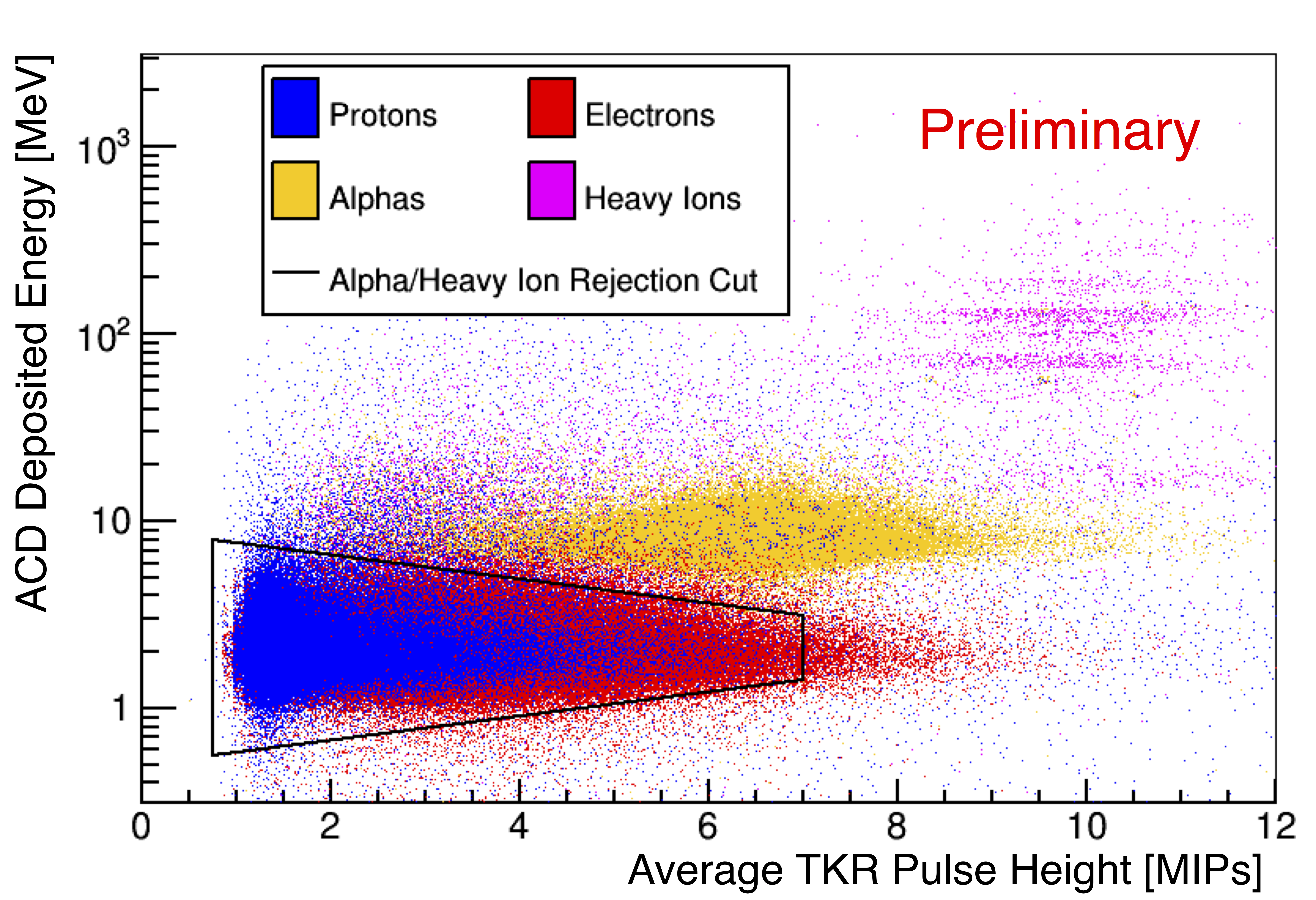}}\quad
	\subfloat[]{\includegraphics[width=7.5cm, height=5cm]{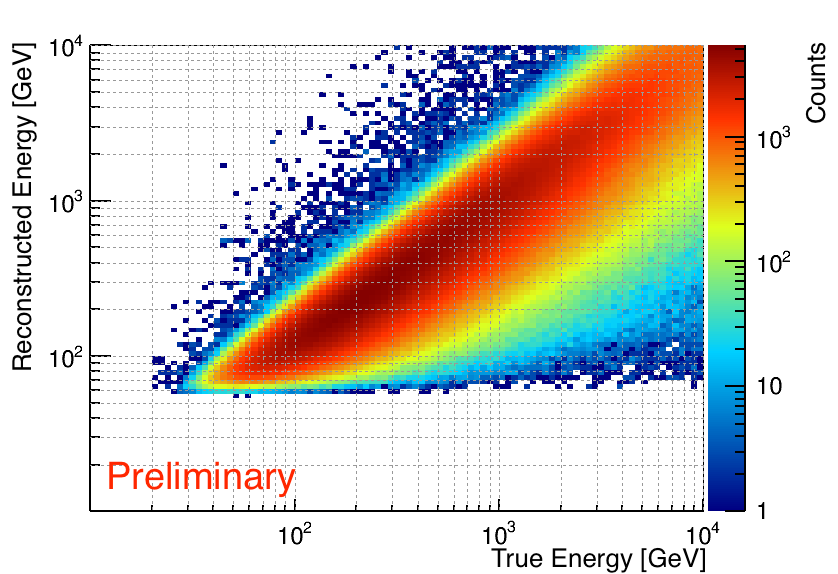}}
 	\caption{\small{Left: Energy deposited in the anti-coincidence detector (ACD) vs the average pulse height in the tracker (TKR), which is used to measure the charge of cosmic rays. Right: Energy response matrix for protons from Geant4 Monte-Carlo simulations.}}
  	\label{alpha_cuts}
\end{figure}

The tracker and anti-coincidence detector are used to remove cosmic rays with charge > 1. The average pulse height in the tracker and charge deposited in the anti-coincidence detector are strongly correlated with the charge of cosmic rays due to the $Z^2$ dependence of ionization loss (Figure \ref{alpha_cuts}). The minimum energy requirement in the ACD also rejects photons to well below the electron flux level~\cite{fermi_cre_spectrum}. We use a classifier developed for the Fermi LAT cosmic-ray electron/positron analyses~\cite{fermi_cre_spectrum} to separate protons from charged leptons. The use of this classifier restricts the energy range of the analysis to 78 GeV - 9.8 TeV in reconstructed energy (i.e. $E_{reco}$). The residual contamination from both heavy ions and leptons is estimated to be < 1\% after applying these cuts. It is worth mentioning that the classifier will also reject residual photons since it classifies events based on the hadronic or leptonic nature of their showers.

The LAT has several observing strategies in addition to standard survey mode. For example, there was modified pointing to study the Galactic center that began in December 2013. In order to remove non-standard observing modes such as this, we require the LAT to be in standard survey mode when each event was recorded. Additionally, we require the rocking angle to be < 52\degree. This removes time periods when the earth is directly in the field of view.

The calorimeter, which is 0.5 interaction lengths at normal incidence, does not fully contain hadronic showers due to its shallow depth. This results in wide energy resolution of protons measured by the LAT, which can be seen in the energy response matrix in Figure \ref{alpha_cuts}. The tail of the energy distribution extends down to tens of GeV, allowing a non-negligible population of low-energy protons into the data set when cutting on $E_{reco}$. These low-energy cosmic rays are deflected by large angles in the geomagnetic field and can create a false positive in an anisotropy search. The influence of the geomagnetic field is greatest towards the earth's horizon. In order to remove cosmic rays arriving from this region, we impose a set of energy-dependent cuts on the LAT's off-axis angle (theta angle of an event relative to the LAT's boresight). This effectively cuts on the field of view, requiring the off-axis angle to be smaller at lower energies. These cuts were developed by performing an analysis in horizontal coordinates (i.e. altitude and azimuth), while remaining blind to the full data set in the equatorial frame. The geomagnetic signal should be maximal in the horizontal frame, while any residual signal is expected to be smeared out in the equatorial frame.

\section{Analysis Methods} \label{methods}
\subsection{Reference Maps} \label{refmaps}
The target sensitivity for this analysis is a dipole amplitude <$10^{-3}$, which is much smaller than the uncertainty in the detector's effective area. Therefore we cannot rely on simulation to estimate the exposure and directly calculate the intensity of cosmic rays from each direction on the sky. The standard method instead is to create a reference map which represents the detector response to an isotropic sky, i.e. the best estimate of what an isotropic sky would look like given the detector's effective area and pointing history. The anisotropy search is then performed by comparing the measured sky map to the reference map.

There are a handful of data-driven methods used to create reference maps. We adopt a time-averaged method similar to direct integration~\cite{milagro_direct_int} and the rate-based method used in~\cite{fermi_cre_anisotropy}. The prescription for the construction of the reference map goes as follows: we first divide the data set into time bins of one year. The use of an integer number of years for the entire analysis removes any contamination from the solar dipole~\cite{icecube_anisotropy}, which is a Compton-Getting-like dipole created by the earth's rotation around the sun. This dipole averages to zero in the equatorial frame over the course of one year. We then calculate the time-averaged rate, $R_{avg}$, in each time bin and construct a PDF of the angular direction of events in the instrument frame, $P(\theta,\phi)$, from the actual distribution of detected events. For each second of live time, the expected number of events is determined by $R_{avg}$ and $P(\theta,\phi)$ gives the probability of detecting an event from any particular direction. Given this information and the instrument pointing history, we then construct the expected sky map for each year. Any underlying anisotropy on the sky is contained in the instantaneous values of $R(t)$ and $P(\theta,\phi,t)$, but averaged out in the time-averaged quantities. This effectively smears the anisotropy over the entire sky in the reference map.

We create 25 independent reference maps in eight energy bins from 78 GeV < $E_{reco}$ < 9.8 TeV and average them in order to beat down statistical fluctuations in the reference map. The maps are then binned cumulatively in energy to create the final energy-integrated maps for the anisotropy search; the results of which are reported as a function of minimum energy.

\subsection{Angular power spectrum} \label{aps}
We perform a spherical harmonic analysis of the relative intensity between the measured sky map and reference map to search for anisotropy. The relative intensity is shown in Equation \ref{relint} where $n_i$ represents data map counts and $\mu_i$ represents reference map counts in the $i^{th}$ pixel.

\begin{equation}
\delta I_i(\alpha_i,\delta_i) = \frac{ n_i(\alpha_i,\delta_i) - \mu_i(\alpha_i,\delta_i)}{\mu_i(\alpha_i,\delta_i)}
\label{relint}
\end{equation}

The relative intensity is decomposed into spherical harmonics (\ref{alm}) and the angular power spectrum (\ref{cl}) is calculated using the anafast algorithm in HEALPix~\cite{healpix}.  In principle, the angular power spectrum is sensitive to anisostropies at all angular scales (angular scale of each multipole $\approx180\degree/l$). We calculate the angular power spectrum up to $l$=30 which corresponds to an angular scale of $\approx6\degree/l$.

\begin{equation}
\hat{a}_{lm} = \frac{4\pi}{N_{pix}} \sum_{i=1}^{N_{pix}} Y^*_{lm}(\pi-\delta_i,\alpha_i)\delta I_i(\alpha_i,\delta_i)
\label{alm}
\end{equation}

\begin{equation}
\hat{C}_l = \frac{1}{2l+1} \sum_{m=-l}^l  |\hat{a}_{lm}^2|
\label{cl}
\end{equation}

In order to determine the significance of the measured power at each multipole, detailed knowledge of the power spectrum under the null hypothesis (isotropic sky) is necessary. This can be calculated directly from the Poisson noise in the map. This white noise level due to finite statistics in the map is given by:

\begin{equation}
C_N = \frac{4 \pi}{N_{pix}^2} \sum_{i=1}^{N_{pix}} (\frac{n_i}{\mu_i^2} + \frac{n_i^2}{\mu_i^3*N_{maps}})
\label{c_n}
\end{equation}
where $N_{maps}$ is the number of independent reference maps that are created and averaged. Equation \ref{c_n} accounts for pixel-to-pixel variations in the sky maps due to non-uniform exposure. With the knowledge of $C_N$, one can calculate the white noise power at each $l$ from the PDF of $C_l$ which follows a $\chi^2_{2l+1}$ distribution~\cite{knox}. Any excess or deficit of the measured angular power compared to the isotropic expectation at a particular multipole indicates an anisotropy at that angular scale. The dipole anisotropy is typically described by the amplitude and phase of a harmonic function. The amplitude of the dipole can be calculated directly from the angular power at $l$=1:

\begin{equation}
\delta=3\sqrt{\frac{C_1}{4\pi}}
\label{delta}
\end{equation}

\section{Results}
\label{results}

\begin{figure}
	\captionsetup[subfigure]{labelformat=empty}
 	\centering
	\subfloat[]{\includegraphics[width=.45\textwidth]{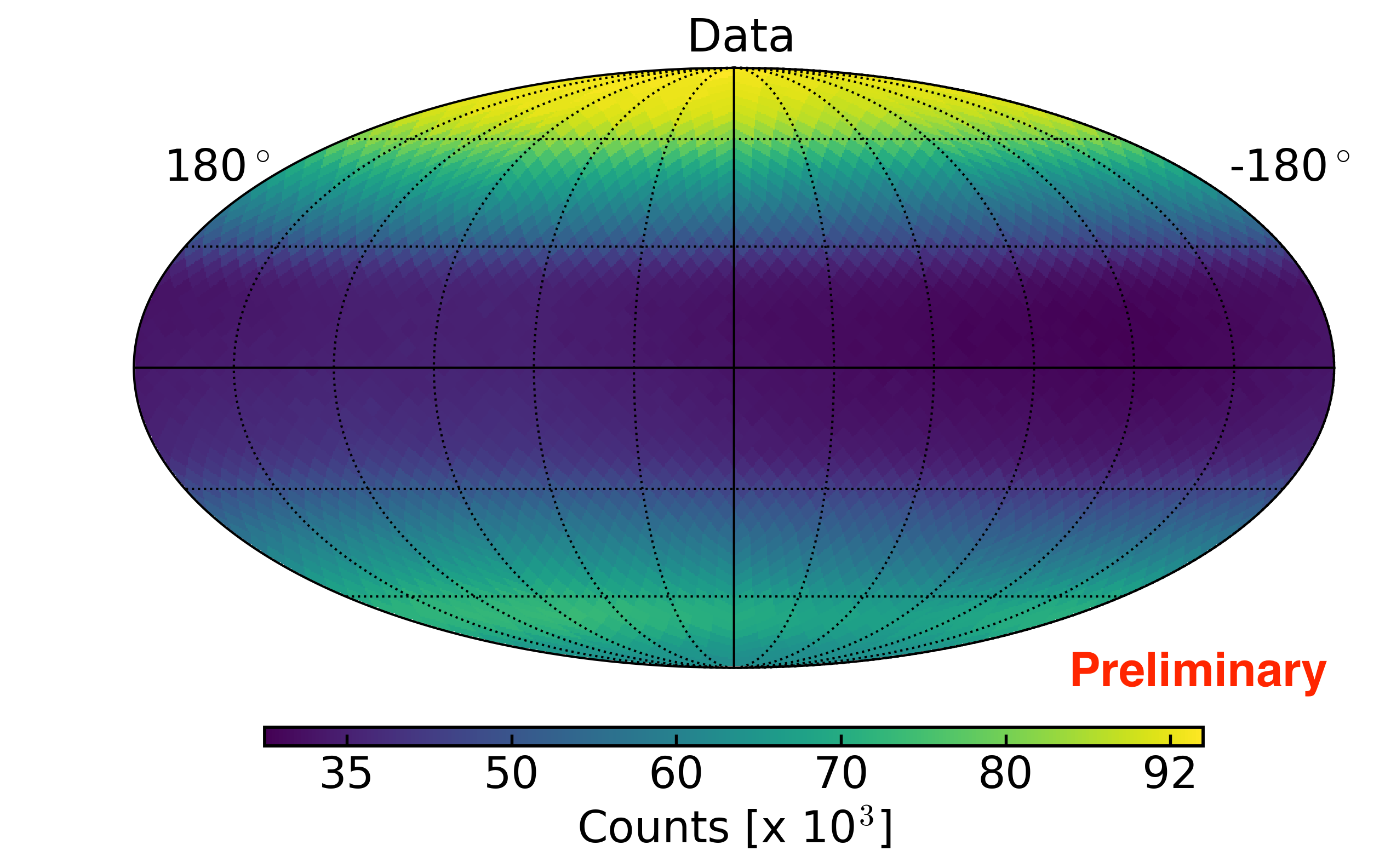}}\quad
	\subfloat[]{\includegraphics[width=.45\textwidth]{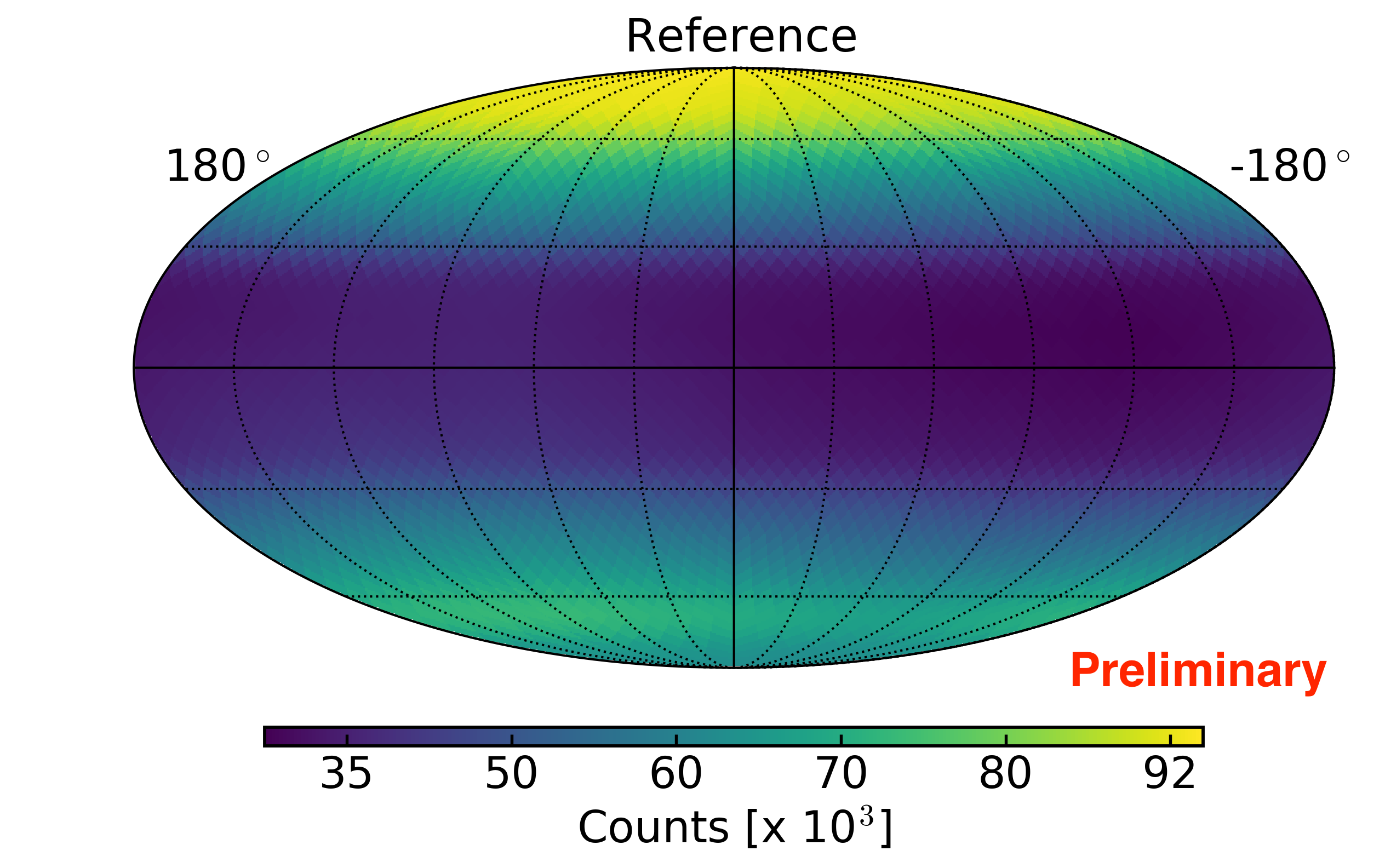}}
 	\caption{\small{Sky maps for events with $E_{reco}$>78 GeV in equatorial coordinates: Data map (left) and the average of 25 realizations of the reference map (right).}}
  	\label{skymaps}
\end{figure}

\begin{figure}
	\captionsetup[subfigure]{labelformat=empty}
 	\centering
	\subfloat[]{\includegraphics[width=.45\textwidth]{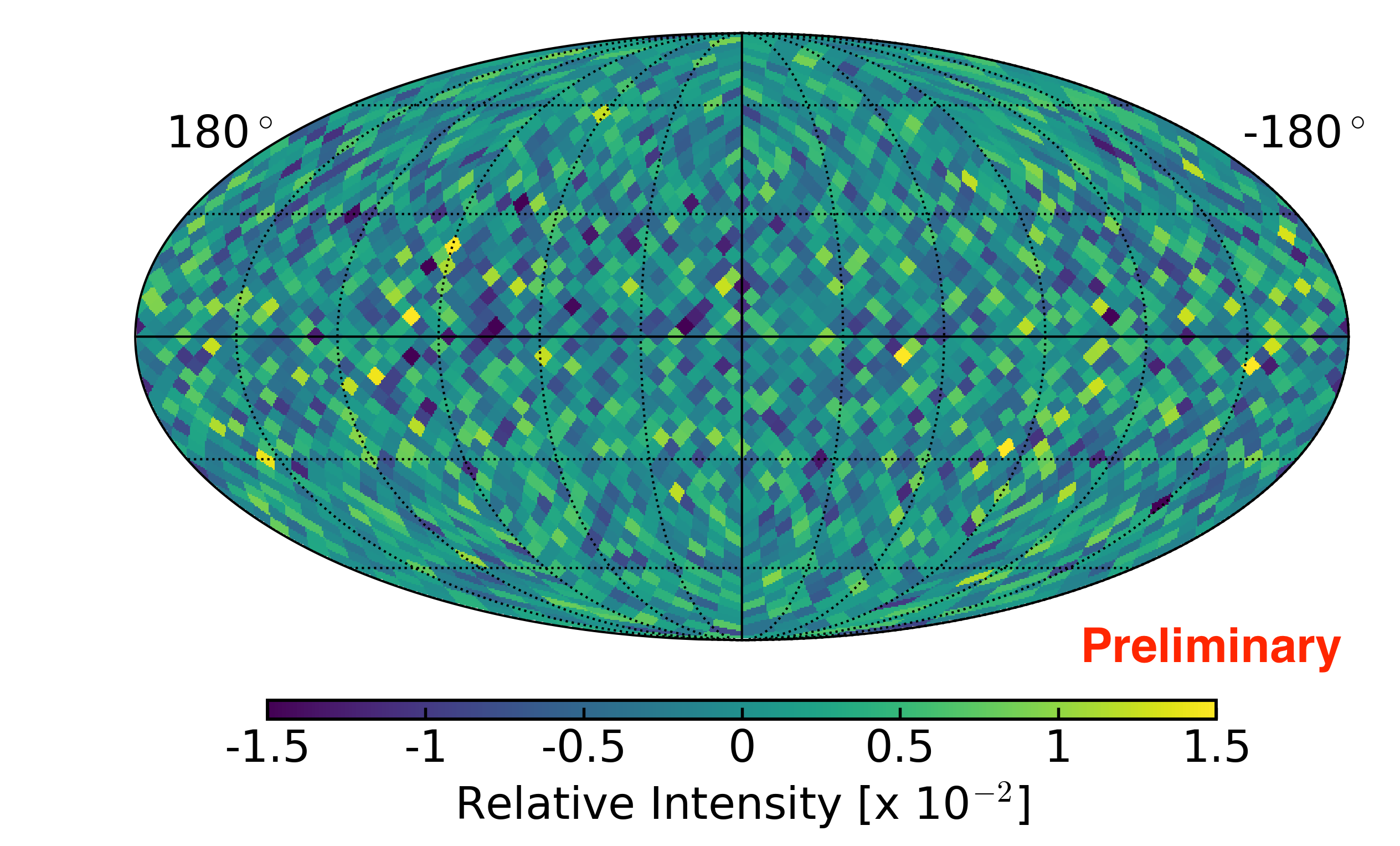}}\quad
	\subfloat[]{\includegraphics[width=.45\textwidth]{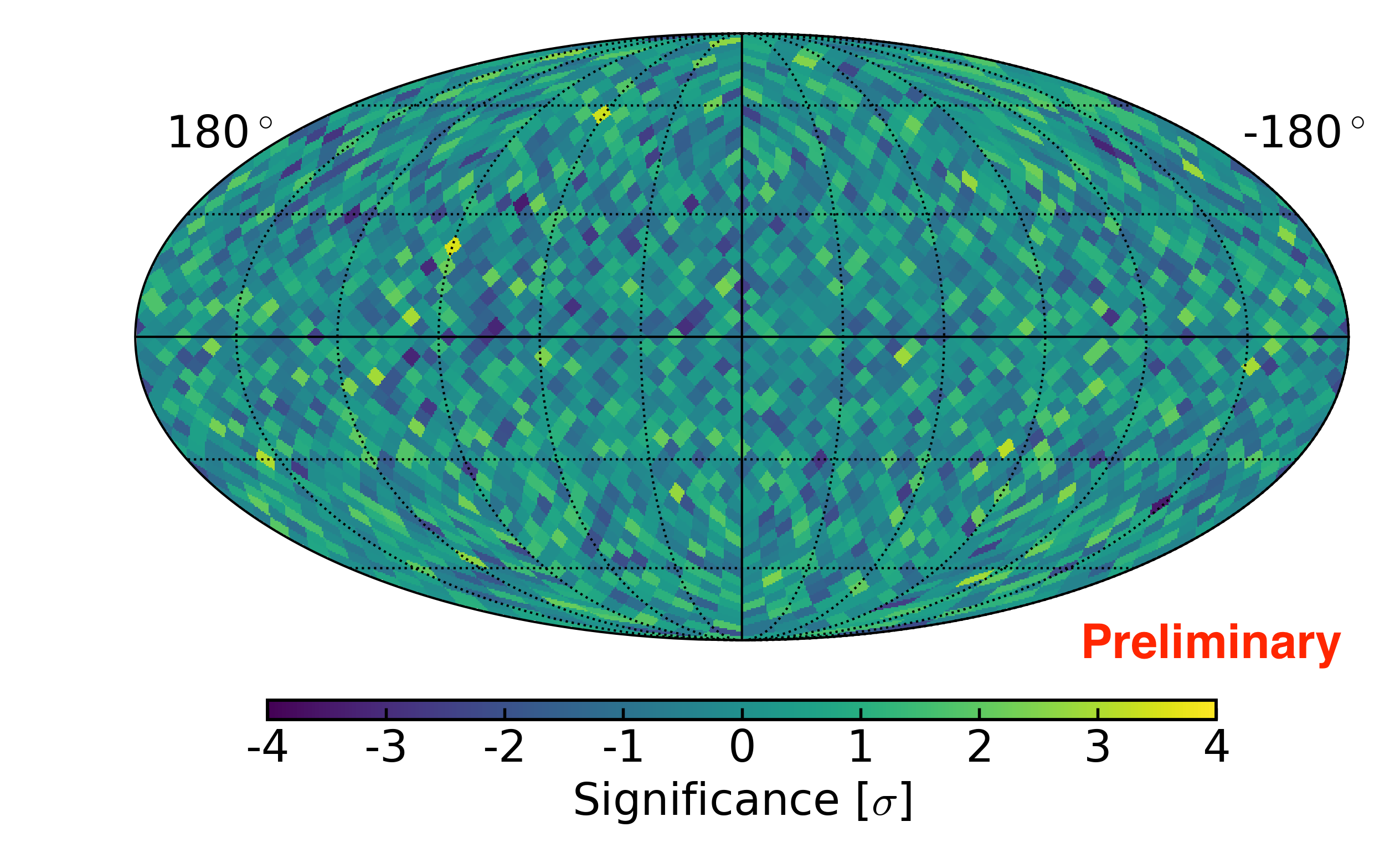}}
 	\caption{\small{Relative intensity and significance maps for events with $E_{reco}$>78 GeV in equatorial coordinates: Relative intensity (left) is given by eq. \ref{relint} and significance (right) is calculated using the Li \& Ma prescription.}}
  	\label{relint_map}
\end{figure}

Sky maps for all events with $E_{reco}$>78 GeV in equatorial coordinates are shown in Figure \ref{skymaps}. The structure seen in these maps is due to the exposure of the instrument. The exposure varies by \textasciitilde60\% across the sky and is greater towards the poles due to the LAT's rocking profile. It is clear to see that the exposure structure in the sky map is maintained in the reference map. Figure \ref{relint_map} shows the relative intensity and Li \& Ma significance maps~\cite{li_and_ma} created from the sky maps in Figure \ref{skymaps}. The relative intensity map exhibits larger fluctuations at equatorial latitudes due to the lower exposure in this region compared to the poles. These pixels are smoothed out in the significance map.

\begin{figure}[h]
	\centering
	\includegraphics[width=.5\linewidth]{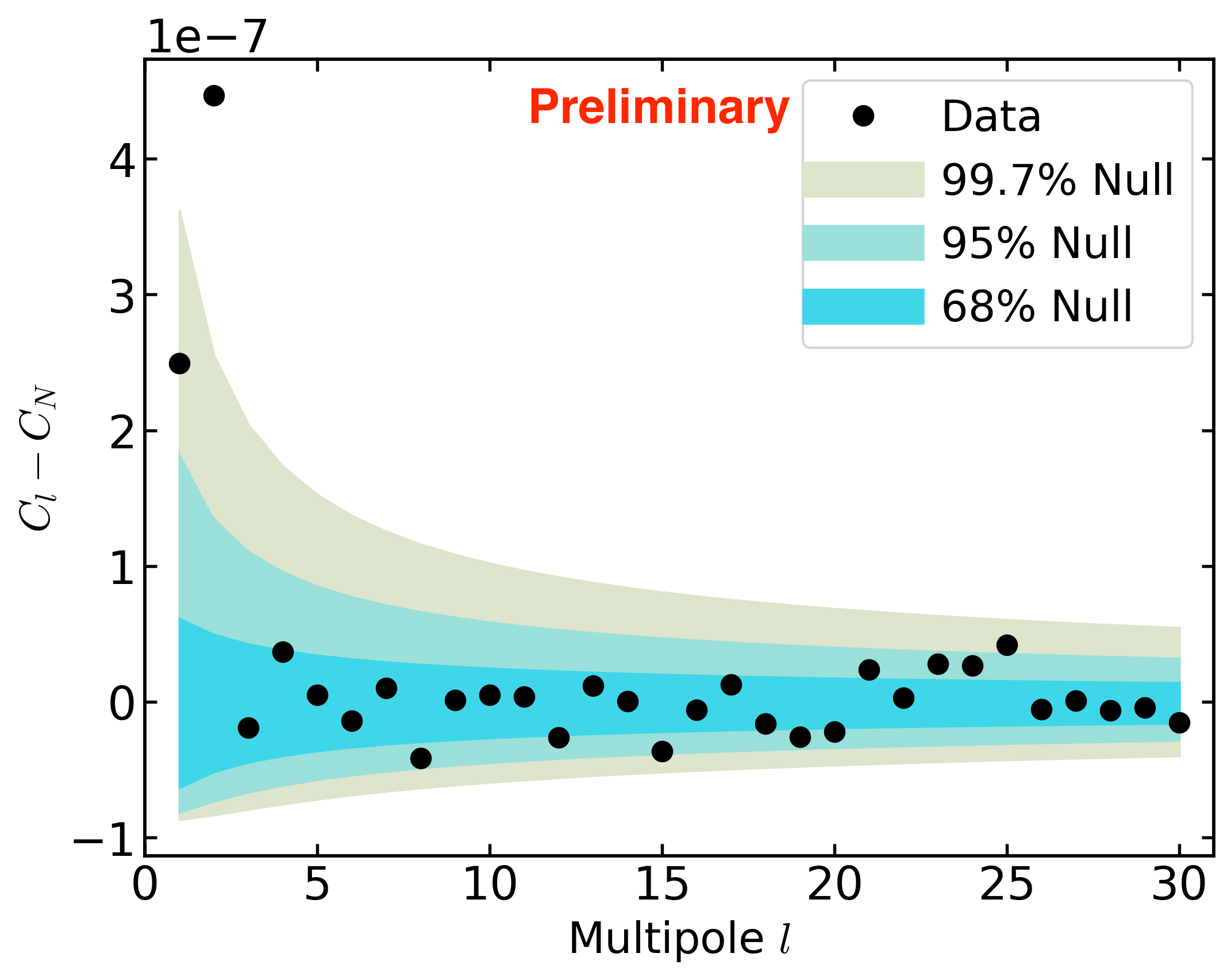}
	\caption{\small{Angular power spectrum for events with $E_{reco}$>78 GeV in equatorial coordinates. The colored bands 			represent the expected distribution of measurements under the null hypothesis (isotropic sky). The source of the significant quadrupole is under investigation.}}
	\hfill
	\label{aps_plot}
\end{figure}

The relative intensity was folded through the angular power spectrum analysis as described in Section \ref{aps}. The angular power spectrum for $E_{reco}$>78 GeV is shown in Figure \ref{aps_plot}. The colored bands represent the 68\%, 95\%, and 99.7\% central regions of expected distribution under the null hypothesis. The bands were calculated directly from the PDF of $C_l$ under the null hypothesis (isotropic sky). The observed power at $l$=1 (dipole) is intriguing at \textasciitilde2.5$\sigma$, but not significant enough to rule out the null hypothesis. The data point at $l$=2 (quadrupole) is significant, but its interpretation is under investigation. Many of the systematics of this analysis exist in the quadrupole due to the LAT's equatorial orbit and further work is being done to rule those out as the explanation of this excess.

\begin{figure}
	\captionsetup[subfigure]{labelformat=empty}
 	\centering
	\subfloat[]{\includegraphics[width=.45\textwidth]{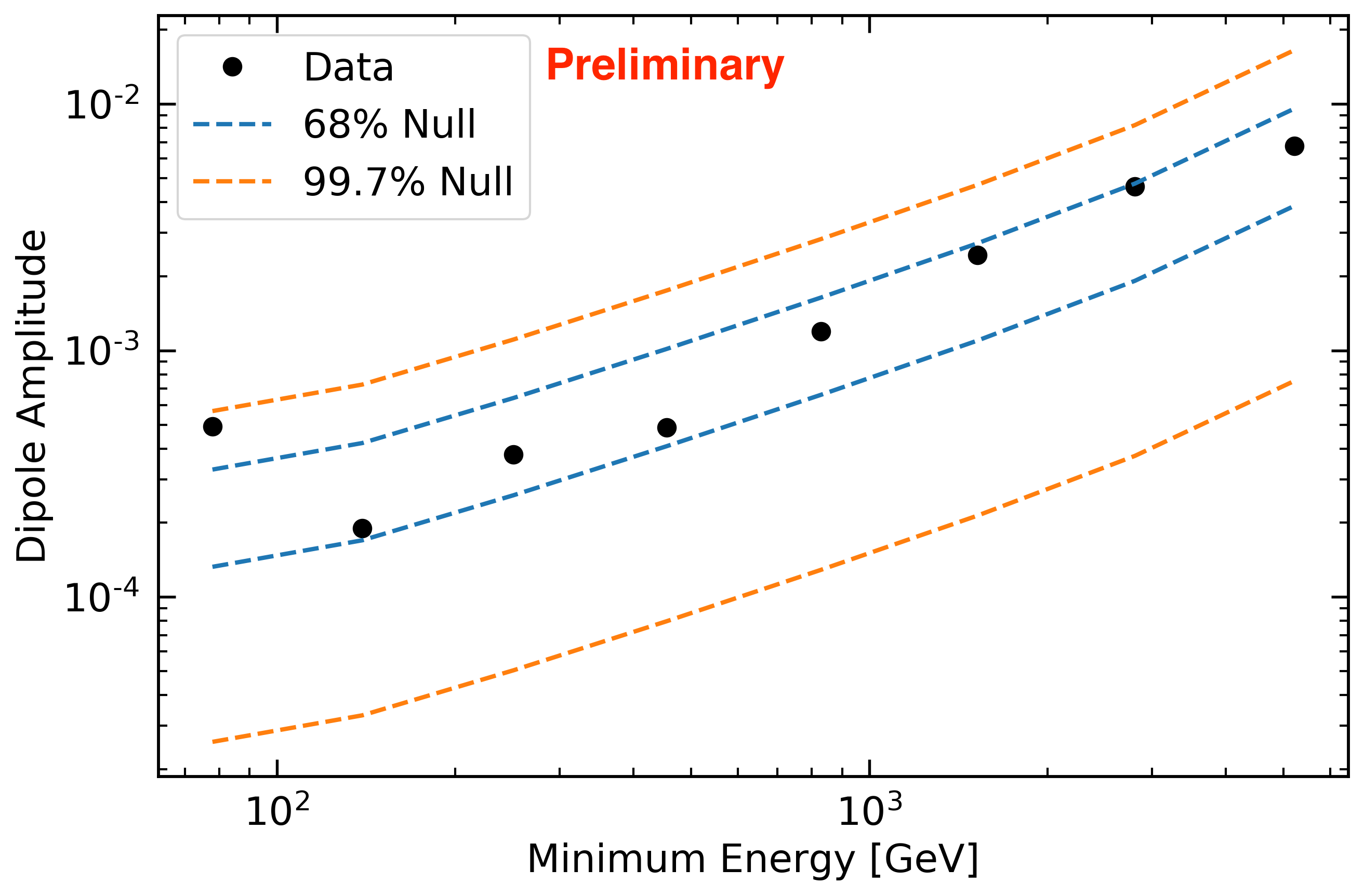}}\quad
	\subfloat[]{\includegraphics[width=.45\textwidth]{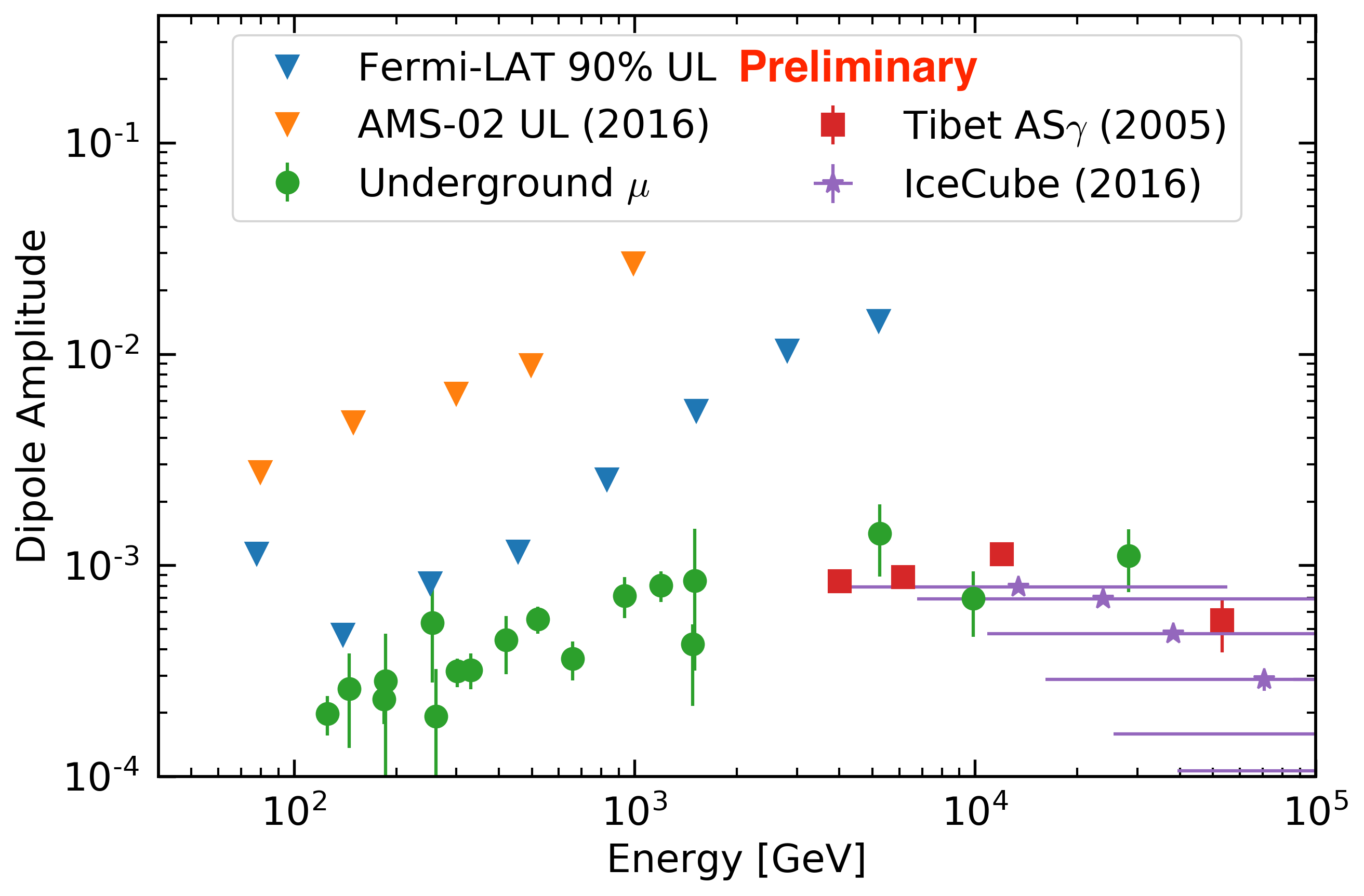}}
 	\caption{\small{Left: Measured dipole amplitude as a function of minimum energy. The 68\% and 99.7\% bands show the expected distribution of measurements for an ensemble of isotropic skies (null hypothesis). Right: Fermi-LAT 90\% upper limits on the dipole amplitude as a function of minimum energy compared to AMS-02 95\% upper limits (minimum energy bins) and ground-based measurements (differential energy bins) from~\cite{AMS,tibet_2005,tibet_2017,icecube_anisotropy}. The AMS-02 analysis is a relative measurement of high-rigidity protons to low-rigidity protons, not an absolute measurement. The ground-based observatories measure the right ascension component of the dipole anisotropy.}}
  	\label{measured_dipole}
\end{figure}

As described in Section \ref{methods}, the analysis is performed on the data set as a function of minimum energy, which yields an angular power spectrum for each minimum energy. We calculate the amplitude of the dipole on the sky from these power spectra using \ref{delta}. Observed dipole amplitudes as a function of minimum energy are plotted in Figure \ref{measured_dipole}. The 68\% and 99.7\% bands show the expected distribution of measurements under the null hypothesis (isotropic sky). All measured amplitudes are consistent with an isotropic sky. Given the non-detection of a significant dipole, we calculate 90\% CL upper limits which can be seen on the right in Figure \ref{measured_dipole}. The upper limits were calculated using the likelihood ratio procedure described in~\cite{fermi_cre_anisotropy}. Since the ground-based measurements in Figure \ref{measured_dipole} are only sensitive to the anisotropy in right ascension, the Fermi LAT upper limits are the strongest to date on the declination dependence of the dipole. 

\section{Conclusion}
\label{conclusion}
We analyzed 160 million cosmic-ray protons detected by the Fermi LAT over the course of eight years and searched for anisotropy in their arrival directions. We did not observe a significant anisotropy at any angular scale except in the quadrupole and further work is underway in order to rule out systematics as its source. We calculated upper limits on the dipole amplitude as a function of minimum energy. Due to the limited reconstruction capabilities of ground-based experiments, these are the strongest limits to date on the declination dependence of the dipole. 

\section{Acknowledgments}
The Fermi-LAT Collaboration acknowledges support for LAT development, operation and data analysis from NASA and DOE (United States), CEA/Irfu and IN2P3/CNRS (France), ASI and INFN (Italy), MEXT, KEK, and JAXA (Japan), and the K.A. Wallenberg Foundation, the Swedish Research Council and the National Space Board (Sweden). Science analysis support in the operations phase from INAF (Italy) and CNES (France) is also gratefully acknowledged.

\bibliography{meehan_ICRC_2017}{}
\bibliographystyle{ICRC}

\end{document}